\documentclass[10pt,conference]{IEEEtran}
\usepackage[utf8]{inputenc}
\usepackage{lscape}
\usepackage{adjustbox}
\usepackage{pdflscape}
\usepackage{geometry}
\usepackage{amsmath}
\usepackage[table,xcdraw]{xcolor}

\title{A literature review on different types of empirically evaluated bug localization approaches}
\author{
\IEEEauthorblockN{F.H. Zamfirov}
\IEEEauthorblockA{\textit{Computer Science and Engineering} \\
\textit{Eindhoven University of Technology}\\
f.h.zamfirov@student.tue.nl}
}

\begin{document}

\renewcommand{\thetable}{\arabic{table}}
\maketitle

\section{Introduction}
Today, software systems have a significant role in various domains among which are healthcare, entertainment, transport and logistics, and many more. It is only natural that with this increasing dependency on software, the number of software systems increases. Additionally, these systems become more and more complex. All this leads to a rise in the number of software faults also known as bugs. As a result, the ability to locate the source of a bug (e.g.\ a file or a commit) is vital for the development and maintenance of efficient software solutions.

Bug localization refers to the automated process of discovering files that contain bugs, based on a bug report. This research project aims to make a literature review on different techniques for bug localization. This study distinguishes itself from other surveys and literature reviews \cite{wong2016survey} in one significant way. The focus of the work is on identifying, categorizing and analyzing existing bug localization methods and tools which were evaluated in an industrial setting. To the best of my knowledge, there are no other works that prioritise this aspect. Unfortunately, such literature is scarce, therefore, bug localization techniques evaluated on open source software are also included.

The structure of the study is as follows. Section \ref{background} provides brief explanations of relevant concepts. Then, Section \ref{overview} contains an overview of the investigated tools and techniques. A more detailed investigation of the various techniques and a recommendation of the most beneficial approach is present in Section \ref{discussion}. Finally, Section \ref{conclusion} concludes this work.

\section{Background}
\label{background}
This section introduces the concepts of bug localization, open source and closed source, as well as some of the common metrics and datasets used in the investigated methods.

\subsection{Bug Localization}
Bug localization, also referred to as fault localization, is the automated process of finding the source of a given software bug based on a bug report. There is a variety of bug localization methods using different strategies and information (i.e.\  data) to locate the origin of a bug. Among the most studied and well-known types are information retrieval based bug localization and spectrum analysis based bug localization.

\subsection{Open source and Closed source}
Open source software projects are publicly available for utilization and modification. Such projects are used in empirical evaluations in the majority of the scientific literature because they are easily accessible. Conversely, closed source projects are proprietary and commonly absent from academic research. Because this study is conducted for an industrial partner, the most relevant tools and techniques are the ones tested in industrial settings and/or on closed source projects. Moreover, the industrial partner is interested in solutions applicable for source code written in C and/or C++, while most open source bug localization datasets contain Java code. Unfortunately, due to their unavailability, studies using closed source projects for evaluations are scarce. Therefore, this study also includes tools evaluated on open source projects.

\subsection{Information retrieval (IR)}
Information retrieval refers to the process of searching for relevant information and/or metadata in a single document or a collection of documents. Web search engines can be considered as an example of IR. A web search is a query for specific information on the web - a collection of documents, and the most relevant websites are returned as top results.

There are many bug localization tools and techniques which are based on IR and share the following generalized strategy. They consider a bug report as a query and a collection of source code files (i.e.\  the project) as a collection of documents. In an attempt to find the source of a bug, the tools find textual overlap between the query (bug report) and the documents (source code files). The source code file(s) with the biggest overlap, are reported in a ranked list, starting with the most relevant one. 

\subsection{Program spectrum}
\label{spectrumb:based}
A program spectrum refers to some specific execution information of a program. For example, this could be information about the execution of conditional statements. A program spectrum can be utilized for tracking program behaviour. 

Spectrum based bug localization is a heuristic involving three elements: a test coverage matrix (1), a hit-spectrum (2) and a fault locator (3). These elements are explained below.

1: All spectrum based bug localization approaches use test coverage information. They collect information about which source code elements are covered by which tests in a test coverage matrix. The columns of the matrix contain the test cases and the rows contain the elements being tested. Each cell in the matrix contains a binary indicator of whether a code element is covered by a test case (value equals one) or not (value equals zero)

2: The hit spectrum is calculated for every element (row) included in the test coverage matrix. The hit spectrum is a tuple with four values $(e_{f},e_{p},n_{f},n_{p})$. In this tuple, $e_{f}$ represents the number of failing test cases that execute the element, $e_{p}$ the number of passing ones. The number of failing and passing test cases that do not cover the element are represented by $n_{f}$ and $n_{p}$ respectively.

3: Finally, an equation known as a fault locator computes the suspiciousness of all tested elements. Sorting all elements by decreasing suspiciousness produces a ranking with the most likely culprit behind a bug at the top of the ranking. The intuition behind the spectrum based bug localization is that when a certain element gets executed more often by failing test cases and less often by passing test cases, it is assigned a higher suspiciousness. Different spectrum based techniques propose their own fault locators, however, there exist some which are widely cited and used \cite{Tarantula,Dstar, ochiai}.

\subsection{Evaluation metrics}
\label{evaluation:metrics}
\textbf{EXAM score:} This score represents the percentage of statements that must be examined before reaching the first statement containing the bug.

$$ \mathrm{EXAM} = \frac{|\textit{\text{ statements examined }}|}{|\textit{\text{ all statements in the program }}|}\cdot 100$$
\textbf{Mean Average Precision (MAP):} This is a widely used IR evaluation metric which considers all buggy files and their corresponding ranks. It is calculated by the following formula:

$$ \mathrm{MAP} = \frac{1}{|Q|}\cdot \mathrm{AP}$$

where $Q$ is the total number of queries and $AP$ is the average precision of a single query, so $MAP$ is the mean of the $AP$ for all queries. $AP$ is defined by

$$ \mathrm{AP} = \sum_{k=1}^{M} \frac{\mathrm{P}(k) \cdot \mathrm{pos(k)}}{\text{\textit{number of positive instances}}}$$

where $M$ is the number of retrieved source files, $P(k)$ is the precision at a cut-off rank  $k$ and $pos(k)$ is a binary value representing whether or not the file at rank $k$ is buggy. $AP$ results in a high value when the buggy files are sorted correctly at the top of the list and results in a low value when they are scattered throughout the ranking.

\textbf{Mean Reciprocal Rank (MRR):} If the source of the bug is at the top in the ranked list of results, its reciprocal rank equals one, if it is the tenth in the list, its reciprocal rank equals $1/10$. Therefore smaller values indicate that the source of a bug is ranked low. Larger values indicate better performance. MRR is the average reciprocal rank over all queries.

$$ \mathrm{MRR} = \frac{1}{|Q|}\sum_{i=1}^{Q} \frac{1}{rank_{i}}$$

\textbf{Recall at Top N:} This metric reports how many bugs with at least one file (i.e.\  the origin of the bug) are present in the top N (N=1,5,10) results that a tool returns. Like MRR, this metric focuses on early precision. The popularity of the metric is based on the assumption that if at least one of the files causing a bug is found early, it may be easier for developers to find the other files.

\subsection{Evaluation datasets}
\label{evaluation:datasets}
\textbf{Defects4j:} Jalali et al.\ \cite{Defects4j} present a dataset containing 357 bugs (initially). These bugs originate from open source projects written in Java. Additionally, the dataset includes a framework with an interface that facilitates conducting and reproducing research. Because the bugs in this dataset are "real-world" bugs (i.e.\  they haven't been engineered on purpose) and the interface is easy to use, Defects4j is widely used in studies on bug localization. Some of the Defects4j projects referenced in this study are Apache Commons Lang, ApacheCommons Math, JFreeChart, Joda-Time and Google ClosureCompiler.

\textbf{iBugs:} Zimmermann et al.\ \cite{iBugs} present the iBugs dataset. This dataset precedes the other two described in this section. It contains a few projects among which, the most notable is AspectJ, which is commonly used for evaluating bug localization projects. 

\textbf{Dataset by Ye et al.\ \cite{ye2014learning}:} In their study Ye et al.\ \cite{ye2014learning} create a benchmark dataset containing six open source projects. Specifically, AspectJ, Birt, Eclipse, JDT, SWT and Tomcat. These projects are used in the original evaluations of some of the more recent works investigated in this study.

\section{Overview of tools and techniques}
\label{overview}
Table \ref{table:1} presents all the methods and tools investigated in this work as well as some important aspects such as their types, what do they localize to (e.g.\ files, statements, commits, etc.), etc. Table \ref{table:2} contains the number of bugs referenced in the original papers that present each of the tools. Interestingly, the numbers vary greatly from a few hundred to tens of thousands. It is likely that this inconsistency results from the difference between the types of approaches. For example, tools and techniques which are Machine Learning and Deep Learning based require a large amount of training data to optimize their performance. The study presenting DNNLoc \cite{DNNLoc}, an approach that uses Deep Learning uses a dataset with 22 747 bugs. For other approaches such as Information Retrieval based ones, this is not needed. Therefore, the number of bugs used during their evaluations is a lot less. Finally, one of the tools, PRFL \cite{PRFL} is evaluated on artificial bugs created through mutation. As mutations are essentially variations of code, they can be easily generated, resulting in a big number of artificial bugs. Nevertheless, there is a trend visible in the ``Type of Bug(s) used for evaluation'' column in Table \ref{table:1}, that researchers investigating bug localization are adopting a preference for real-world data (bugs).
\newpage
\newgeometry{left=0.5cm,bottom=0.5cm}
\begin{landscape}
\begin{table}[hbt]
\caption{Tools and techniques considered in this study}
\label{table:1}
\begin{adjustbox}{width=1.52\textwidth}
\begin{tabular}{|l|l|l|l|l|l|l|l|l|}
\hline
\multicolumn{1}{|c|}{\textbf{Technique/Tool}} &
  \multicolumn{1}{c|}{\textbf{Type}} &
  \multicolumn{1}{c|}{\textbf{\begin{tabular}[c]{@{}c@{}}Localize\\ To\end{tabular}}} &
  \multicolumn{1}{c|}{\textbf{\begin{tabular}[c]{@{}c@{}}Implementation \\ Availability\end{tabular}}} &
  \multicolumn{1}{c|}{\textbf{\begin{tabular}[c]{@{}c@{}}Type of Bug(s) \\ used for evaluation\end{tabular}}} &
  \multicolumn{1}{c|}{\textbf{\begin{tabular}[c]{@{}c@{}}Evaluation \\ metrics\end{tabular}}} &
  \multicolumn{1}{c|}{\textbf{Evaluation dataset}} &
  \multicolumn{1}{c|}{\textbf{Advantages}} &
  \multicolumn{1}{c|}{\textbf{Disadvantages}} \\ \hline
BLUiR \cite{BLUiR} (2013) &
  IR &
  file &
  unavailable &
  real-world bugs &
  \begin{tabular}[c]{@{}l@{}}- MAP\\ - MRR\\ - Recall at Top N\end{tabular} &
  \begin{tabular}[c]{@{}l@{}}Bugs from four open source projects \\ (AspectJ, Eclipse, JDT, ZXing)\end{tabular} &
  \begin{tabular}[c]{@{}l@{}}- Considers structure of source code (classes, methods, \\ variables, comments) to increase performance (structured \\ information retrieval)\\ - Can utilize bug similarity data to increase performance\end{tabular} &
  \begin{tabular}[c]{@{}l@{}}- Relies on programming constructs from object-oriented\\ languages\\ - Has a runtime overhead\\ - Performance relies on good naming conventions and well\\ written bug reports.\end{tabular} \\ \hline
Bug2Commit \cite{B2C} (2021) &
  \begin{tabular}[c]{@{}l@{}}IR,\\ Machine\\ Learning\end{tabular} &
  commit &
  unavailable &
  \cellcolor[HTML]{FFFFFF}{\color[HTML]{000000} \begin{tabular}[c]{@{}l@{}}- client-side crashes \\ (mobile app)\\ - serverside performance \\ regressions\\ - mobile simulation tests \\ for performance\end{tabular}} &
  \begin{tabular}[c]{@{}l@{}}- MRR\\ - Recall at Top N\end{tabular} &
  \begin{tabular}[c]{@{}l@{}}400 crashes\\ 40 perf. regressions\\ 550 regression tests\end{tabular} &
  \begin{tabular}[c]{@{}l@{}}- Can handle complex bug reports containing \\   various features (e.g.\ summary, stack trace, etc.)\\ - Can handle synonyms using weighted word \\   embedding\end{tabular} &
  \begin{tabular}[c]{@{}l@{}}- Result is dependent on overlap in relevant words \\   between a bug report and bug source artefact.\\ - Not possible to weigh words from \\   different features differently.\\ - Uses corpus to weigh words, but not domain \\   knowledge.\\ - Mispredictions may potentially waste time and \\   resources.\end{tabular} \\ \hline
CBT \cite{CBT} (2012)  &
  \begin{tabular}[c]{@{}l@{}}Statistical\\ Debugging\end{tabular} &
  \begin{tabular}[c]{@{}l@{}}executable \\ statement\end{tabular} &
  unavailable &
  \begin{tabular}[c]{@{}l@{}}unspecified,\\ (some artificially injected \\ bugs)\end{tabular} &
  - EXAM Score &
  \begin{tabular}[c]{@{}l@{}}22 programs (both C and Java)\\ (The Siemens suite, the Unix suite,\\ space, grep, gzip, make and Ant\end{tabular} &
  \begin{tabular}[c]{@{}l@{}}- Only requires coverage information\\ - Outperforms the compared (then state-of-the-art) Tarantula \\ technique\\ - Evaluated on a large set of projects of different sizes and \\ programming languages\end{tabular} &
  \begin{tabular}[c]{@{}l@{}}- Evaluated on a single metric\\ - Some of the bugs in the projects are artificial and were\\ seeded by the researchers.\end{tabular} \\ \hline
\begin{tabular}[c]{@{}l@{}}Convolutional Neural Network\\ \cite {CNN} (2019)\end{tabular} &
  \begin{tabular}[c]{@{}l@{}}Deep\\ Learning,\\ IR\end{tabular} &
  file &
  \begin{tabular}[c]{@{}l@{}}description for\\ re-implementation\\ available\end{tabular} &
  real-world bugs &
  \begin{tabular}[c]{@{}l@{}}- AUC\\ - MAP\\ - MRR\\ - Recall at Top N\end{tabular} &
  \begin{tabular}[c]{@{}l@{}}Bugs from five open source projects\\ (AspectJ, Eclipse, JDT, SWT, Tomcat)\end{tabular} &
  \begin{tabular}[c]{@{}l@{}}- Outperforms IR based bug localization techniques.\\ - Is able to address issues related to language semantics\\ (e.g.\ synonymy)\\ - Efficient prediction time\end{tabular} &
  \begin{tabular}[c]{@{}l@{}}- High training time.\\ - Requires exessive of hardware resources - GPUs and\\   memory.\\ - Models are difficult to use by mainstream software\\    practitioners\end{tabular} \\ \hline
CooBa \cite{CooBa} (2021) &
  \begin{tabular}[c]{@{}l@{}}Adversarial\\ Transfer\\ Learning\end{tabular} &
  file &
  unavailable &
  real-world bugs &
  \begin{tabular}[c]{@{}l@{}}- MAP\\ - Recall at Top 10\end{tabular} &
  \begin{tabular}[c]{@{}l@{}}Bugs from four open source projects\\ (AspectJ, Eclipse, JDT, SWT)\end{tabular} &
  \begin{tabular}[c]{@{}l@{}}- Leverages the existance of projects with rich historical bug data\\ - Negates the transfer of private information of one project to another\\  (which impacts performance)\end{tabular} &
  \begin{tabular}[c]{@{}l@{}}- Tool evaluated only on source and target projects \\ written in the same programming language\\ - Sizes of projects and their effects on performance are \\ not documented.\end{tabular} \\ \hline
D\&C \cite {D&C} (2019) &
\begin{tabular}[c]{@{}l@{}}IR,\\ Machine\\ Learning\end{tabular} &
  file &
  available &
  real-world bugs &
  \begin{tabular}[c]{@{}l@{}}- MAP\\ - MRR\\ - Recall at Top N\end{tabular} &
  Bench4BL &
  \begin{tabular}[c]{@{}l@{}}- Significantly outperforms all state-of-the art IR tools on MAP and\\ MRR\\ - Adaptively calculates the most effective weights applied to the\\ similarity scores of IR features of a given pair of bug report and \\ source code file\end{tabular} &
  \begin{tabular}[c]{@{}l@{}}- Requires extensive hardware resources\\ - Requires a way to split the used dataset for multi-classification\end{tabular} \\ \hline
DNNLoc \cite {DNNLoc} (2017) &
  \begin{tabular}[c]{@{}l@{}}Deep \\ Learning,\\ IR\end{tabular} &
  file &
  unavailable &
  real-world bugs &
  \begin{tabular}[c]{@{}l@{}}- MAP\\ - MRR\\ - Recall at Top N\end{tabular} &
  \begin{tabular}[c]{@{}l@{}}Bugs from six open source projects\\ (AspecJ, Birt, Eclipse, JDT, SWT, \\ Tomcat\end{tabular} &
  \begin{tabular}[c]{@{}l@{}}- Reasonable prediction time.\\ - Is able to link bug reports and source code that do not contain \\ similar wording\\ - Higher accuracy than other state-of-the-art\\  machine-learning approaches\end{tabular} &
  \begin{tabular}[c]{@{}l@{}}- High training time\\ - Requires extensive hardware resources\end{tabular} \\ \hline
  Legion \cite{Legion} (2021) &
  \begin{tabular}[c]{@{}l@{}}IR,\\ Machine\\ Learning\end{tabular} &
  file &
  unavailable &
  \begin{tabular}[c]{@{}l@{}}consumer-reported \\ issues\end{tabular} &
  \begin{tabular}[c]{@{}l@{}}- Developer \\ expectations\\ - MAP\\ - MRR\\ - Recall at Top N\end{tabular} &
  \begin{tabular}[c]{@{}l@{}}Seven of the most functionally\\ important repositories in \\ Adobe Analytics\\ product.\end{tabular} &
  \begin{tabular}[c]{@{}l@{}}- Can utilize bug similarity data to increase performance.\\ - Is parameterised to increase performance.\\ - Correctly identifies a faulty file in the Top 10 recommendations\\  at least 70 \% of the time during evaluation. (junior developer\\ expectation)\end{tabular} &
  \begin{tabular}[c]{@{}l@{}}- Training of the model may be computationally \\   expensive for big repositories.\\ - Is not able to correctly identify a faulty file in \\ the top 5 recommendations at least 80\% of the time\\ during evaluation. (senior developer expectation)\end{tabular} \\ \hline
  \begin{tabular}[c]{@{}l@{}}Patterned Spectrum Analysis\\ \cite{ItemSetMining} (2016)\end{tabular} &
  \begin{tabular}[c]{@{}l@{}}Patterned\\ Spectrum \\ Analysis\end{tabular} &
  method &
  unavailable &
  real-world bugs &
  \begin{tabular}[c]{@{}l@{}}- Wasted effort\\ - Recall at Top 10\end{tabular} &
  \begin{tabular}[c]{@{}l@{}}Bugs from five open source projects\\ (Apache Commons Lang, Apache\\ Commons Math, JFreeChart,\\ Joda-Time, Google Closure \\ Compiler)\end{tabular} &
  \begin{tabular}[c]{@{}l@{}}-  Technique is highly relevant to bugs that rarely occur, but have a \\ significant impact such as those in integration tests\\ - Outperforms state-of-the-art raw spectrum analysis bug localization\\ techniques (in wasted effort)\end{tabular} &
  \begin{tabular}[c]{@{}l@{}}- Unable to rank methods which do not contain method \\ calls.\\ - Less than half of all bugs ranked in  Top 10\end{tabular} \\ \hline
\begin{tabular}[c]{@{}l@{}}PredFL \cite{PredFL} (2019)\end{tabular} &
  \begin{tabular}[c]{@{}l@{}}Spectrum\\ Analysis, \\ Statistical \\ Debugging\end{tabular} &
  \begin{tabular}[c]{@{}l@{}}executable \\ statement\end{tabular} 
   &
   available
   &
   real-world bugs
   &
   \begin{tabular}[c]{@{}l@{}}- EXAM Score\\ - Recall at Top N\end{tabular}
   &
   \begin{tabular}[c]{@{}l@{}}Bugs from five open source projects\\ (Apache Commons Lang, Apache\\ Commons Math, JFreeChart,\\ Joda-Time, Google Closure \\ Compiler)\end{tabular}
   &
   \begin{tabular}[c]{@{}l@{}}- Combines two types of bug localization\\ - Can be used to further improve the performance of an existing \\ bug localization technique\end{tabular}
   &
   \begin{tabular}[c]{@{}l@{}}- Is evaluated as a complementary approach to\\ another bug localization technique (i.e. not standalone)\\ - Implementation works only on Java, as it relies on Java\\ Development Tools (JDT)\\ - Statistical Debugging requires a way to seed predicates\\ into the program\end{tabular}
   \\ \hline
   PRFL \cite{PRFL} (2017) &
  \begin{tabular}[c]{@{}l@{}}Spectrum\\ Analysis \end{tabular} &
  method &
  available &
  \begin{tabular}[c]{@{}l@{}}real world bugs,\\ artificial bugs\end{tabular} &
  \begin{tabular}[c]{@{}l@{}}- Absolute\\ wasted effort\\ - Recall at Top N\end{tabular} &
  \begin{tabular}[c]{@{}l@{}}Bugs from five open source projects\\ (Apache Commons Lang, Apache\\ Commons Math, JFreeChart,\\ Joda-Time, Google Closure \\ Compiler)\end{tabular} &
  \begin{tabular}[c]{@{}l@{}}-  Boosts the performance of existing Spectrum based \\ techniques\\ - Added overhead is insignificant \\ (order of seconds)\end{tabular} & \begin{tabular}[c]{@{}l@{}}- Limited performance on multi-faults\\ - Not investigated whether performance\\ increase is sufficient in practice\end{tabular} \\ \hline
\end{tabular}
\end{adjustbox}
\end{table}
\end{landscape}
\pagebreak
\restoregeometry

\begin{table}[h]
\caption{Number of bugs used during evaluation of the tool/technique}
\label{table:2}
\centering
\begin{tabular}{|l|c|}
\hline
\multicolumn{1}{|c|}{\textbf{Technique/Tool}}                                               & \textbf{Nr of bugs}                                                                                       \\ \hline
BLUiR \cite{BLUiR}                                                                          & 3 479                                                                                                     \\ \hline
\begin{tabular}[c]{@{}l@{}}Bug2Commit\\ \cite{B2C}\end{tabular}                             & \begin{tabular}[c]{@{}c@{}}400 (crashes),\\ 40 (perf. regressions),\\ 550 (regressing tests)\end{tabular} \\ \hline
CBT \cite{CBT}                                                                              & 420                                                                                                       \\ \hline
\begin{tabular}[c]{@{}l@{}}Convolutional \\ Neural Network\\ \cite{CNN}\end{tabular}        & 17 331                                                                                                    \\ \hline
CooBa \cite{CooBa}                                                                          & 17 513                                                                                                    \\ \hline
D\&C \cite {D&C}                                                                            & 5 321                                                                                                     \\ \hline
DNNLoc \cite{DNNLoc}                                                                         & 22 747                                                                                                    \\ \hline
Legion \cite{Legion}                                                                        & 933                                                                                                       \\ \hline
\begin{tabular}[c]{@{}l@{}}Patterned \\ Spectrum Analysis \\ \cite{ItemSetMining}\end{tabular} & 357                                                                                                       \\ \hline
PredFL \cite{PredFL}                                                                        & 357                                                                                                       \\ \hline
PRFL \cite{PRFL}                                                                        & \begin{tabular}[c]{@{}l@{}}357 (real world bugs)\\ 30 692 (artificial bugs)\end{tabular}                                                                                                     \\ \hline
\end{tabular}
\end{table}

\section{Discussion}
\label{discussion}
This section contains more detailed information about the tools and techniques presented in Table \ref{table:1} in Section \ref{overview}. Each approach is introduced, followed by a brief overview of its methodology. However, as some of the approaches are based on very specific knowledge (e.g.\ deep learning) and for the sake the brevity, the methodology is briefly reported with a lot of details omitted. For the full details please refer to the original papers. In this section the terms ``technique'', ``method'' are used interchangeably.

\subsection{BLUiR \cite{BLUiR}}
Saha et al.\ \cite{BLUiR} observe that although IR is widely used for bug localization, a lot of the tools and techniques consider the source of information (bug reports, source code) as texts without any structure. They argue that the structure of source code (methods, comments, variables, etc.) and that of bug reports (title, description, etc.) can be taken advantage of to boost the performance of IR based bug localization. For example, if there exists a class named ``C'' with four variables containing ``C'', the class name doesn't have a high impact and this file could be easily overlooked if there is a file with more than five occurrences of the term. However, the relevance of the class named "C" should be higher than others just containing the term in some variable names. Therefore they propose BLUiR \cite{BLUiR}, an approach that considers both the structure of queries (i.e.\  bug reports) and documents (i.e.\  source code) to enhance the performance of IR in bug localization. Although Saha et al.\ \cite{BLUiR} do not share their implementation publicly, they use the open source Indri toolkit \cite{indri} for their retrieval model and explain in detail how their implementation works. Therefore, I believe it can be reproduced. Additionally, they do provide a link to the dataset they use for their experiments. Unfortunately, the link does not work.

\subsubsection{Methodology}
BLUiR's process begins with building an abstract syntax tree (AST) of every source code file. This tree is traversed to collect names of methods, classes, variables and comments. All of these are tokenized and passed to Indri for stopword removal, stemming and indexing. Two improvements over previous methods are also proposed. First, unlike other IR techniques, programming keywords are not pruned. Saha et al.\ \cite{BLUiR} argue that words such as "String" might occur in some names and pruning them will decrease recall. Instead, by using the extraction of identifiers from the AST, programming keywords are excluded while keeping the ones contained in identifiers. Subsequently, identifiers are split into tokens using techniques such as camel case splitting. However, the full identifiers are kept and indexed as well. This is done as exact identifiers may be present in bug reports. In particular, a bug report may be written by knowledgeable developers or there could be a stack trace attached to the bug report. Saha et al.\ \cite{BLUiR} note that this small change results in a big improvement. To account for the structure of source code and bug reports, Saha et al.\ \cite{BLUiR} distinguish four source code fields (class, method, variable and comments) and two bug report fields (summary and description). They perform a search in each of the eight combinations (bug report field, source code field) and then sum the scores of all combinations. The benefit of this is that when a term appears in multiple fields in a file, such as class name and methods, greater importance is implicitly assigned to the file because of the summation of scores.

\subsubsection{Evaluation}
In order to evaluate the performance of BLUiR, Saha et al.\ \cite{BLUiR} use the same dataset and evaluation metrics used for the then state-of-the-art BugLocator \cite{BL}. The dataset consists of four open source projects (Eclipse, SWT, AspectJ, ZXing). The evaluation metrics for the experiments are Recall at Top N, MRR and MAP (cf.\ Section \ref{evaluation:metrics}). Saha et al.\ \cite{BLUiR} conduct several comparisons between both their approach and BugLocator, as BugLocator outperformed many other techniques at the time and could utilize bug similarity data to further boost its performance, as is explained next.

Initially, the results of BLUiR are compared with those of BugLocator without the use of bug similarity data. It is observed that on three out of four projects, BLUiR outperforms BugLocator by a great margin on all metrics. Subsequently, the results of BLUiR are compared against the improved results of BugLocator utilizing bug similarity data. Even in this case, BLUiR performs better. The authors decide to modify BLUiR to be able to use bug similarity data as well. However, this does not boost performance as much as it did for BugLocator \cite{BL}.   Saha et al.\ \cite{BLUiR} argue that this is because BLUiR is able to compensate for the lack of bug similarity data. Nevertheless, BLUiR can take advantage of such data whenever it is available.

\subsubsection{Advantages and Disadvantages}
The biggest benefit of BLUiR is that it takes into consideration the structure of source code and bug reports and does not treat them as simple text. Nevertheless, this makes the approach dependent on the use of good naming conventions and well-written bug reports. In a context in which both of these are missing, the performance may significantly decline. Moreover, the approach as described by Saha et al.\ \cite{BLUiR} uses object-oriented (OO) programming constructs such as class names etc. for the aforementioned fields. Therefore this approach is language dependent. Although the experiments in the study are done on Java code, the authors note that BLUiR should be easily adaptable to other OO languages. 

Finally, as structured information retrieval requires more computations there is a runtime overhead. This overhead depends on the size of the source code collection and it may vary between 3x and 12x. Even so, the full execution time is in seconds, meaning that the overhead may be negligible.

\subsection{Bug2Commit \cite{B2C}}
Murali et al.\ \cite{B2C} propose an information retrieval (IR) based tool which is meant to address practical concerns around industrially used IR based bug localization. The goal of the tool is to identify a bug introducing commit, based on a bug report and a list of candidate commits (gathered using time or build information from the bug report). The tool was proposed and evaluated at Facebook.

\subsubsection{Methodology}
There are several requirements at Facebook that a bug localization technique must fulfil. Firstly, it must localize to a commit and not to a file. Secondly, the method has to be unsupervised. Although Facebook contains a lot of historical data on bug fixes, there is no data on bug causes. This would result in a lack of labelled training data. Thirdly, the method must be capable to process complex queries and documents. Bug reports and code commits at Facebook contain several different components, referred to as features, which may vary. For example, a bug report may have various stack traces from different threads, metadata, exception message etc. Code commits also could have different features such as comments, summary, test plan, etc. Therefore, these entities cannot be simply thought of as individual collections of words. Finally, because Facebook utilizes a monolithic repository that receives thousands of commits daily containing code for several platforms, in numerous languages and conventions, the technique needs to handle word similarities and the idiosyncratic coding conventions.

Based on these requirements Murali et al.\ \cite{B2C} investigate several existing bug localization methods of which only two are aligned with the requirements - Orca \cite{bhagwan2018orca} and Locus \cite{wen2016locus}. These techniques are evaluated and deemed insufficient. Consequently, the researchers propose Bug2Commit, an tool using IR and semantic word embedding to address the aforementioned requirements.

\subsubsection{Evaluation}
The performance of Bug2Commit is assessed on three Facebook datasets - one containing mobile app crashes, one containing server performance regressions and one containing mobile simulation tests. Recall at Top at 1, 5 and 10 and MRR are used as evaluation metrics (cf.\  Section \ref{evaluation:metrics}). When compared against Orca \cite{bhagwan2018orca} and Locus \cite{wen2016locus}, Bug2Commit performs better or at least as well.

When analysing the three cases individually Murali et al.\ \cite{B2C} observe that the tool is beneficial for server-side performance regressions and mobile app crashes. In the former, the tool reduces the time spent on localizing from hours (time without the use of the tool) to minutes. In the latter, for a period of 6 months, Bug2Commit successfully localizes four out of seven bugs. Among the three unsuccessful cases, two contain corrupted data and one is miss-predicted. Finally, the mobile simulation tests differ from the other two cases as the consumer of the localization is a system rather than a human. In this case, it is observed that synonymous words are used within the query and candidates. Like the previous techniques (i.e.\  Orca, Locus) Bug2Commit is unable to localize efficiently. Thus, the researchers use Bug2Commit with the weighted word embedding model. This version of the tool is able to capture word semantics well.

\subsubsection{Advantages and Disadvantages}
One benefit of Bug2Commit is that it can be applied in scenarios in which the sources of information are complex entities. Both bug reports and the code commits may contain multiple components. Additionally, due to the weighted word embedding, the tool is able to grasp the semantics of words.

Nevertheless, Word overlap between the origin of a bug (i.e.\  a commit) and the bug report is needed for Bug2Commit to work correctly. Applied in cases where there is insufficient overlap, the performance of the tool may decrease drastically. Another disadvantage is the lack to prioritize words from a feature (e.g.\ bug report summary, bug report title, etc.). In some scenarios, bug titles may contain more relevant information than other features. However, only word frequency is used in calculating the word weight, while domain knowledge is not. Finally, mispredictions may lead to resource and time waste, therefore caution needs to be applied.

 \subsection{CBT \cite{CBT}}
Crosstab based technique (CBT) \cite{CBT} is a statistical technique that utilizes code coverage information on the level of executable statements as well as the outcome of all test cases (i.e.\  whether a test passed or failed) to localize bugs. The name of the technique comes from the use of a table depicting the relationship between two or more variables referred to as a crosstab. Such a table is computed for every (executable) statement and a statistic is used to calculate the statement's suspiciousness. The main difference between CBT  and heuristic based techniques (such as some of the spectrum based techniques) is the use of well-defined statistical analysis which is not present in heuristic based techniques.
 
\subsubsection{Methodology}
As previously stated, crosstabs are used for investigating the relationship between two or more variables. Wong et al.\ \cite{CBT} construct crosstabs for each executable statement. Each crosstab contains two column-wise variables \textit{covered} and \textit{not covered} and two row-wise variables \textit{successful} and \textit{failed}. For every crosstab, a hypothesis test is made in order to check for a dependence relationship between the coverage of a statement and the result of the program execution. The chi-square test is used when deciding whether to accept or reject the hypothesis that the coverage of the statement and the execution result are independent. Laghari et al.\ \cite{CBT} note that the degree of association between the coverage of a statement and the execution result is more interesting than the (in)dependence relationship itself. To that end, several fractions such as the number of failed tests over the number of all tests help in determining whether the coverage is associated more with the failed or the successful execution result. Using all the gathered information, statements are then separated into five classes. Four of these classes contain statements with a (high or low) degree of association between their coverage and the (failed or successful) execution result. The fifth class contains statements with independent coverage and execution result. The statements from the first class (i.e.\  high association degree between coverage and failed execution result) are presumably the most likely for containing bugs.

\subsubsection{Evaluation}
Wong et al.\ \cite{CBT} perform an extensive evaluation by investigating several aspects. The main evaluation metric used through the experiments is the \textit{EXAM score} (cf.\  Section \ref{evaluation:metrics}).

Firstly, the performance of CBT  is evaluated against that of Tarantula \cite{Tarantula}, a well-known spectrum based technique. The performances of both these approaches are recorded on both large and small projects (i.e.in terms of \textit{lines of code} (LOC)). The small projects (i.e.\  less than 1000 LOC) used are a part of the Siemens and Unix suites which have been used in many fault localization studies (before 2014). Both suites consist of multiple programs, all written in C. Because both CBT and Tarantula output ranked lists in which several statements could have the same suspiciousness value, Wong et al.\ \cite{CBT} consider two cases per approach. The first case is referred to as the best case, in which a statement containing a bug is examined before statements that do not contain a bug and have the same rank. The second case is the exact opposite. Therefore it is referred to as the worst case. In experiments done on the Siemens dataset, not only CBT's best case is better than that of Tarantula, but its worst case is also better than Tarantula's worst. The two techniques are compared on larger projects as well. Namely, on the space program, grep, gzip, and make. However, the size of the projects does not impact the results of the comparison. CBT seems to outperform Tarantula, except for two test cases.

Secondly, CBT  is evaluated against two other statistical approaches. Namely, Sober\cite{liu2005sober} and Liblit05 \cite{Liblit}. This evaluation contains both a qualitative and a quantitative aspect. In the quantitative evaluation, CBT always produces better results than Liblit05 generally outperforms Sober.

In addition, Wong et al.\ \cite{CBT} also investigate the effectiveness of CBT in another language - Java, as all previously mentioned programs are in C. For that, a comparison between CBT and Tarantula is performed on the software project Apache Ant. In this comparison the difference between the performance of both approaches is significant. specifically, the worst performance of CBT is better than the best of Tarantula.

\subsubsection{Advantages and Disadvantages}
Even though the evaluation done by Wong et al.\ \cite{CBT} presents CBT as a very efficient technique there are several considerations to be made. Firstly, Wong et al.\ \cite{CBT} utilize mutation based fault injection to generate artificial bugs. Therefore, not all bugs used for the experiments can be considered real. Secondly, although CBT significantly outperforms Tarantula, at the current point in time of writing this study, there are many newer works on spectrum based bug localization proposing better performing than Tarantula methods.

\subsection{Convolutional Neural Network \cite{CNN}}
Deep Learning (DL) \cite{deng2014deep} has been employed in a variety of fields and has gained popularity due to its boosted performance over that of general Machine Learning (ML) approaches in areas such as image processing \cite{krizhevsky2012imagenet}, speech recognition \cite{hinton2012deep} and others. Deep learning based models can be used for bug localization as well, as linking bug reports and source code can be transformed into a classification problem. Moreover, Deep learning can address one of the big challenges of bug localization  - the lexical gap between bug reports and source code. More specifically, the use of synonymous words in bug reports and source code files. Polisetty et al.\ \cite{CNN} investigate the benefits for developers of applying deep learning based models for bug localization by evaluating the performance of a Convolutional Neural Network (CNN) and Simple Logistic Model against that of several existing bug localization techniques. 

As a motivation for the study Polisetty et al.\ \cite{CNN} cite a survey \cite{kochhar2016practitioners} in which 386 software practitioners across 5 continents provide their expectations of bug localization research. One of the insights from the survey is that even the tools and approaches with the best performance cannot satisfy 75\% of the participants. Therefore, the authors explore whether state-of-the-art Deep Learning based models satisfy the expectations of practitioners and how do these models perform in comparison to other bug localization techniques.

\subsubsection{Methodology}
To evaluate the two models (CNN and Simple Logistic Model) the researchers extract the source code for each open source project and using publicly available bug reports and bug-commit mappings, label each file as \textit{buggy} or \textit{/non-buggy} based on whether the file is changed in a bug fixing commit. For each project, Polisetty et al.\ \cite{CNN} construct three datasets of source code. One containing all files in a project, one containing just buggy files and one containing very buggy files (i.e.\  files linked to multiple bugs). This is done to investigate the effect of varying buggy files on performance. After extracting the data, it is pre-processed by removing numbers, punctuation symbols, programming-related keywords. Stemming and the application of other pre-processing schemes used in other DL related studies is done as well. Then a traceability matrix is constructed, which is the Cartesian product of the list of all source files and bug reports. For every record in this matrix, the corpus of the bug report and that of the source code are merged. With this, the data is prepared for training and testing the models. For each dataset, 90\% of the data is used for training and validation and 10\% is used for testing.

\subsubsection{Evaluation}
Polisetty et al.\ \cite{CNN} evaluate the performance of multiple models on five open source bug localization datasets - AspectJ, Tomcat, SWT, Eclipse, JDT. Two of the models are implemented and evaluated directly and the others are accessed through the means of publicly available statistics. The two implemented models are a CNN with an architecture proposed by Kim \cite{kim2014convolutional} and a traditional ML technique - Simple Logistic model \cite{landwehr2005logistic}.

When comparing the performance of the CNN to that of the Simple Logistic Model, the researchers conclude that the former outperforms the latter in most cases, but has higher training time. Additionally, training Simple Logistics models is faster but they require a big amount of memory (in the magnitude of terabytes). Moreover, the computational resources, needed for training both the DL and ML bug localizing models, are both expensive and difficult to obtain.
 
When comparing the CNN model to another CNN \cite{huo2017enhancing}, the implementation of Polisetty et al.\ \cite{CNN} performs worse. However, they argue that this is because the other implementation is run on a reduced set of source files, which impacts performance and therefore can also be misleading to software practitioners. 

Finally, comparing the performance of the CNN to that of IR techniques, such as BugLocator\cite{BL} and BLUiR \cite{BLUiR}, the CNN outperforms the other techniques in every metric.

Lastly, the researchers discuss the effect of varying buggy files on performance. They note that the CNN model is considerably better performing on smaller datasets. Coincidentally, most DL based models consider only a subset of source code files when evaluating performance. Therefore, this overly-optimistic performance evaluation can be misleading to developers.
 
\subsubsection{Advantages and Disadvantages}
The CNN that Polisetty et al.\ \cite{CNN} evaluate outperforms the widely cited and best performing IR based techniques. This may be because the CNN addresses the challenge with which IR-techniques struggle. Namely, the use of synonymous words in source code and bug reports. Additionally, the model has an efficient prediction time (4 seconds per 1000 files). However, these benefits come with a trade off. Firstly, high training time. While the model can handle projects with millions of lines of code the training time would be in the magnitude of months. An excessive amount of hardware resources is also needed (e.g.\ GPUs and memory). Polisetty et al.\ \cite{CNN} further note that such a type of technique is difficult to use by mainstream software practitioners. 

\subsection{CooBa \cite{CooBa}}
Zhu et al.\ \cite{CooBa} observe that although there is a multitude of supervised machine learning based bug localization techniques, which show promising results, these techniques require rich historical data for training purposes. They argue that this may not always be achievable, such as in the case of a newly developed project. There exist cross-project bug localization techniques which are based on the idea of transferring knowledge from a project, rich with historical data (source project), to a project lacking this data (target project). However, these techniques fail to capture the uniqueness of a project. As a result, the performance of such methods is negatively impacted. Therefore, Zhu et al.\ \cite{CooBa} present an approach based on adversarial learning called Cross-project Bug Localization via Adversarial Transfer Learning (CooBa) which outperforms other state-of-the-art cross-project bug localization techniques.

\subsubsection{Methodology}
Initially, a set of bug reports and a collection of source code files for the source project (i.e.\  the project with historical bug data) and the same sets for the target project (i.e.\  the project which will leverage this data) are collected. In addition, two indicator matrices are constructed per project. These matrices indicate whether a file is buggy or not w.r.t a bug report. The cross-project bug localization is initiated as a classification learning task. During the training of the framework, the goal is to learn prediction functions by input pairs (bug report, source code file) from both the source and target projects. After training the model, the prediction function for the target project is used to compute the relationship of each pair of a bug report and source code file in the target project. There are three integral parts in the model - shared bug report processing, cooperative code file processing and relevance prediction. The cooperative code file processing contains four components, one of which is the public feature extraction. It is in this component where the adversarial training is used to guarantee the effective learning of public features. The adversarial training methodology \cite{goodfellow2014explaining} usually refers to the use of a generator and a discriminator which are considered as adversaries. In CooBa there is a shared feature extractor that acts as a generator and a discriminator that tries to distinguish from which project (source or target) is a code file from.

\subsubsection{Evaluation}
The performance of CooBa is compared against that of both other cross-project bug localization methods, and techniques that work within a single project, refereed to by Zhu et al.\ \cite{CooBa} as \textit{within-project} methods. Among these methods, two are investigated in this study. Namely, BugLocator \cite{BL} and DNNLoc \cite{DNNLoc}. The evaluation dataset consists of four open source projects - AspectJ, Eclipse, JDT, SWT. For the evaluation, one of the projects is used as a source project and another is used as a target project (e.g.\ source: AspectJ, target: Eclipse). All combinations of such pairs are considered. The evaluation results in two main observations. First, \textit{within-project} techniques trained on one (source) project do not perform well when directly used on another (target) project. This indicates the need for special cross-project bug localization techniques. Secondly, CooBa \cite{CooBa} outperforms other cross-project techniques w.r.t both MAP and Recall at Top 10 (cf.\  Section \ref{evaluation:metrics}).

In addition, Zhu et al.\ \cite{CooBa} evaluate the benefits of the adversarial transfer learning. They do so by changing the model to CooBa* which does not utilize adversarial transfer learning and compare its performance to that of CooBa on all project pair combinations using the MAP metric. CooBa outperforms CooBa* which indicates that without the use of adversarial transfer learning, noise from one project is brought with the transfer of knowledge. This noise negatively impacts bug localization.

\subsubsection{Advantages and Disadvantages}
The ability to utilize the historical bug data of one project in order to perform bug localization on another is CooBa's biggest advantage. However, Zhu et al.\ \cite{CooBa} only consider the scenario in which both projects are written in the same programming language. In addition, the study lacks an analysis of how the size of the source project and the amount of historical bug data impacts performance and is there an optimal size and/or amount of bug data.

\subsection{D\&C \cite{D&C}}
Koyuncu et al.\ \cite{D&C} who propose the Divide and Conquer method (D\&C) motivate their study on the observations of a previous study by Lee et al.\ \cite{lee2018bench4bl}. In particular, the observations that although newer and better IR based bug localization tools and techniques are constantly being introduced, they (generally) do not get adopted by software practitioners. Primarily, this is due to the limited performance of the state-of-the-art methods and the absence of detailed validation on the importance of different IR features. Koyuncu et al.\ \cite{D&C} investigate six state-of-the-art IR based bug localization methods (BugLocator, Brtracer, BLUiR, Amalgram, Blia and Locus) and observe that a significant portion of bugs gets localized exclusively by each of the tools. They argue that different tools are more appropriate for specific sets of bugs reports. In addition, they find additional connections between the performance of the tools and specific sets of IR features. This prompts Koyuncu et al.\ \cite{D&C} to develop an approach in which the weight of similarity scores between the features of the source code and those of the bug report are learned for different groups of bug reports. That is the Divide and Conquer method (D\&C) \cite{D&C}.

\subsubsection{Methodology}
In order to adaptively compute the most efficient weights for the similarity scores of IR features, Koyuncu et al.\ \cite{D&C} decide to use a supervised learning technique in which examples from the dataset are used to learn a classification. Furthermore, a multi-classifier approach is chosen. This approach consists of building and training several classifiers, each one trained on a different part of the dataset. Instead of choosing one classifier at the end, the output of all of them is combined by taking the average of the prediction probabilities.

In their investigation of the state-of-the-art tools, Koyuncu et al.\ \cite{D&C} observe that there exist sets of bug reports that are only localized by a certain tool and that there is a set successfully localized by all the tools. This observation and the one that the difference between the tools is primarily in the IR features that are considered leads Koyuncu et al.\ \cite{D&C} to separate the dataset into regions based on the performance of the tools. This is done to provide relevant data for the various classifiers. For example, there is a region for each state-of-the-art tool containing the bug reports on which the tool performed best. Classifiers trained on these regions are referred to as region-specific classifiers. There are other regions as well, such as the one containing data on which all the tools performed well.  Subsequently, for each region, the data is separated into sets for validation and training. However, the data for each region is very imbalanced as a small part of the pairs of bug reports and source files are buggy. The classification algorithm LightGBM, which can account for data imbalances, is used to address the issue. 

Each classifier is trained in 10 000 iterations in combination with early stopping, an approach used to avoid overfitting. The training stops when there is no further improvement and the best model is found. Subsequently, the best model of each classifier is used on the relevant parts of the data to obtain the probability values for each combination of a bug report and source file. 

A combined output ranking is made by averaging the probabilities for source code files in each model.

\subsubsection{Evaluation}
D\&C is evaluated on the Bench4BL which at the time contained data from 46 projects written in Java. Originally the dataset contained 8 652 reports, however, for the evaluation of D\&C it was cleaned from bug reports linked to missing files or ones assumed to be post-fix activities. In particular, these are bug reports having the same person as reporter and fixer.
The evaluation metrics are the widely used MAP, MRR and Recall at Top 1,5,10 (cf.\  Section \ref{evaluation:metrics}).

Initially, D\&C is compared against each of the studied state-of-the-art tools individually. From this comparison, Koyuncu et al.\ \cite{D&C} observe that D\&C is able to localize more bugs than all state-of-the-art tools at Top 1, Top 5 and Top 10 with up to 13\% at Top 1. In addition, there is an improvement in MAP and MRR as well. Then, another comparison is made on the project level. Meaning that the performance of the state-of-the-art tools is compared against those of D\&C for each individual project in the dataset (having D\&C trained on the entire dataset). In these experiments, the MAP and MRR performance vary significantly, with some dropping very low. Nevertheless, D\&C outperforms other tools for the majority of projects. Finally, the impact of the multi-classification is evaluated by analysing the performance of specific classifiers. There are several interesting observations. Firstly, D\&C outperforms the region-specific classifiers and the classifier trained on bug reports which were localized at Top 1 at least by one tool. This supports the claim that D\&C finds an efficient way to compute the most effective weights for the similarity scores of IR features. Secondly, the classifier trained on the data for which no tool was successful at Top 1, performs better than some of the region-specific classifiers. Koyuncu et al.\ \cite{D&C} take this as a confirmation for the need to divide the dataset.

\subsubsection{Advantages and Disadvantages}
In the study of Koyuncu et al.\ \cite{D&C} D\&C has been trained on heterogeneous data  (i.e.\  from different projects). However, the investigation of the performance after training on a single project is scarce. Koyuncu et al.\ \cite{D&C} do train D\&C on the data from the biggest project in the dataset, but do not report specific statistics about its performance. Furthermore, the impact of the number of bugs reports is not mentioned, in all of the experiments D\&C is trained on the entire dataset. While D\&C is able to outperform several widely cited IR based bug localization the lack of the above data needs to be considered. Finally, its practicality is also questionable. Koyuncu et al.\ \cite{D&C} split the dataset based on the performance of other state-of-the-art tools. This might not be optimal or feasible for software practitioners.

\subsection{DNNLoc \cite{DNNLoc}}
DNNLoc \cite{DNNLoc} is a method that is designed to alleviate the already mentioned challenge for IR based bug localization - the use of synonymous language, also referred to as lexical mismatch, by Lam et al.\ \cite{DNNLoc}. It does so by using deep neural networks (DNNs) which learn to connect words from a bug report to different terms in source code files.

\subsubsection{Methodology}
The model is trained by creating two types of pairs between source code and bug reports. The first type is the positive pair and it is between a bug report and a file containing the bug cause. The second type, the negative pair, is between a bug report and files containing textual similarity to the bug report, however, not the bug cause. For each pair, several features are collected and feature vectors are constructed. Features are collected in the following way.

The bug reports are parsed and preprocessed using traditional IR techniques such as stopword removal and stemming. Identifiers mentioned in the bug report are not removed. Instead, they are split into words that are also stemmed. However, the full identifiers are kept as well. After the prepossessing Term Frequency - Inverse Document Frequency, a traditional IR technique, is used to calculate the significance weights of the words. These weights are the features of the bug report.

The source code files contain four features of interest. Identifiers, names of API classes and interfaces, comments and textual descriptions of API methods and classes. These are all extracted and processed similarly as bug reports.

The textual similarity between a bug report and a source code file is also considered. It is calculated using the rVSM model utilized by Bug2Commit \cite{B2C} as well. Lam et al.\ \cite{DNNLoc} note that the textual similarity and the relevancy feature (computed at a later point) will suit each other and contribute to linking a bug report and source files, both containing and lacking similar words.

Bug fixing and metadata features are collected as well. Among these is a score representing how recent a file has been fixed for a bug, and a score for the number of times a file has been fixed before a certain bug report. Lam et al.\ \cite{DNNLoc} argue that such metadata has been previously shown to improve performance.

Finally, three DNN models are employed. The first one is an autoencoder which reduces the dimensions of the feature vectors. The benefit of doing so is twofold. On one hand, it reduces computational costs and makes the approach more scalable. On the other hand, redundant information is removed in the process. The output of the autoencoder is used as input to the second DNN - a relevancy estimator. This DNN produces a score on how relevant a source file is to a bug report. The last DNN takes all previously mentioned features and computes a single score which indicates how relevant a source file is to a bug report. Such scores are calculated for all files and ranked. 

\subsubsection{Evaluation}

Lam et al.\ \cite{DNNLoc} conduct a series of experiments on a dataset provided in a previous study containing six open source projects. They use three widely employed metrics. Namely, Recall at Top N, MAP, and MRR (cf.\  Section \ref{evaluation:metrics})

Initially, the importance of the different features (e.g.\ relevancy feature, metadata etc.) is tested. On its own, the DNN computed relevancy is very inaccurate. However, combining it with the metadata feature significantly increases its accuracy. The textual similarity is more accurate than both previously mentioned components. However, combining it with the relevancy and metadata features improve its results by 2,5-8\%. Furthermore, Lam et al.\ \cite{DNNLoc} note that in 18 cases neither the textual similarity nor the relevancy score, rank the correct file in the top 20, however, when combined (i.e.\  a component considering both) they do.

Subsequently, the performance of the DNN is tested. Specifically, accuracy with varying sizes of the DNN Estimator (i.e.\  number of nodes), accuracy with different training data sizes and the ability to link terms in source files with words in bug reports. Lam et al.\ \cite{DNNLoc} observe that the DNN is able to recognize connections between semantically connected words. For example, the DNN correctly identifies that the word context is relevant to the terms "ctx" and "envCtx".

Lam et al.\ \cite{DNNLoc} do also compare the results of DNNLoc against those of several state-of-the-art bug localization techniques. One based on IR - BugLocator \cite{BL}, one based on ML - Naive Bayes and a hybrid one, a Learn to Rank method by Ye et al.\ \cite{ye2014learning}. In comparison to BugLocator, DNNLoc achieves significant improvement on Top 1 and 5 accuracy. Lam et al.\ \cite{DNNLoc} argue that on one hand, the DNN model addresses the lexical gap challenge. On the other hand, the metadata features additionally enhance accuracy. DNNLoc outperforms the Naive Bayes and the Learn to rank approach as well due to some inherent weakness of those methods.

\subsubsection{Advantages and Disadvantages}
DNNLoc shows that deep learning can indeed improve the results achievable with IR and can address the lack of word overlap between bug reports and source files. In addition, although rSVM is used in the study by Lam et al.\ \cite{DNNLoc}, these results may be improved even further by experimenting with another IR method. Unfortunately, as with other deep learning bug localization techniques, there is a computational cost. Lam et al.\ \cite{DNNLoc} do not explicitly mention the time efficiency of DNNLoc apart from stating that training time is large. They do state that this can possibly be alleviated by parallel computing for DNN, however, this is not investigated. The prediction time for a single bug report is within minutes.

\subsection{Legion \cite{Legion}}
Legion \cite{Legion} is an extended implementation of the well-known bug localization approach based on information retrieval (IR) BugLocator (BL), proposed by Zhou et al.\ \cite{BL}. Jarman et al.\ \cite{Legion} who propose Legion are initially interested in investigating the merit of BugLocator for developers in Adobe Analytics. Specifically, they are interested in whether the tool can adhere to developer expectations defined for two groups of developers. For the first group - developers new to a repository, the solution needs to correctly identify a faulty file in the top 10 recommendations at least 70\% of the time. For the second group consisting of developers familiar with a repository, the solution needs to correctly identify a faulty file in the top 5 recommendations at least 80\% of the time.

\subsubsection{Methodology}
As in Adobe Analytics, Jira is used for issue tracking and Git is used for source control, the following process is employed in preparation for the BugLocator evaluation. In every repository, all commit messages are scanned for a Jira issue id relating to a bug. Whenever one is found, information about it is collected from both Git and Jira, such as commit message and id, bug description, bug summary, reporter, etc. To avoid data-set biases reported by prior bug localization studies, the researchers consider customer-reported bugs because formal procedures are applied for their handling.
 
The performance of BugLocator is evaluated on 933 bugs, using the same evaluation metrics as those originally used by Zhou et al.\ \cite{BL} - MAP, MRR and Top 1,5,10. In comparison to the evaluation results in the original paper (i.e.\  experiments on 4 open source projects), the results on the repositories at Adobe are worse. Because these results do not meet the developer expectations, the researchers decide to extend BugLocator with additional corpora and configurable parameters. BL+ is the result of this extension.

BL+ has four configurable parameters, one for pre-processing (2 options) and three for the computation of a \textit{rSVMScore} (63 options), a \textit{SimiScore} (2 options) and a \textit{FinalScore} (11 options). However, these additional parameters and all their possible combinations result in 2,772 possible configurations. For each bug, it is observed that a different configuration may perform best, but without a way of knowing the optimal configuration apriori, the researchers decide to use all configurations. They do so by computing the final score for each configuration as well as the sum of scores, referred to as a \textit{stacked score}. The stacked score and individual scores are then used as features along labels indicating whether a file is buggy, to learn a supervised model using Random Forest, which scores the features to optimize results. Legion is the name of this final solution containing BL+ and the supervised model.

\subsubsection{Evaluation}
For the evaluation, seven Adobe Analytics repositories of high importance are chosen. In terms of functionality, these repositories cover user interface components, microservices, backend collections, etc. Six of the seven repositories contain between 300 - 700 source code files and one is considerably bigger with the number of files ranging between 2500 and 5000.

When comparing the three implementations (i.e.\  BL, BL+ and Legion), BL+ boosts the results of BugLocator by 29.0\%, 14.1\% and 8.7\% in terms of Top 1, 5, and 10. Legion improves the same scores by 143.4\%, 58.1\% and 36.4\%, respectively. In terms of time efficiency Legion takes 15.12 minutes on average for training and testing of the Random Forests for six repositories. For the seventh, largest repository, running all the 2,772 BL+ parameter configurations takes 34.8 minutes on average. This leads to the conclusion that the time efficiency of applying the bug localization technique on a bug report is highly dependent on the size of the corpora and the number of files in a repository. Furthermore, the researchers argue that Legion's time efficiency can be considered from two perspectives. The first perspective is the time it takes to train the model, which is of little concern for Adobe Analytics as training may be run as a background process periodically, while older versions of the model handle incoming bug reports. The second, and more important perspective, is the time it takes to run the trained model. At the time of the study, the process of getting a bug in the hands of the right developer took days, this provides ample time for Legion to run and construct a list of potential buggy files. The researchers conclude that BugLocator can be useful in an industrial setting and that with some augmentation it is able to fulfil one developer expectation, that of developers new to a repository.

\subsubsection{Advantages and Disadvantages}
While Legion has the ability to leverage bug similarity data due to being an extension to BugLocator, its main benefit comes from having adjustable parameters. In combination with supervised learning, an optimal configuration of parameters can be chosen, providing the best results. Additionally, Jarman et al.\ \cite{Legion} note that at least 70\% of the time the tool correctly identifies a faulty file in the Top 10 recommendations. Nevertheless, supervised learning is computationally expensive, especially for big repositories.

\subsection{Patterned Spectrum Analysis \cite{ItemSetMining}} 
Laghari et al.\ \cite{ItemSetMining} present the patterned spectrum analysis as a bug localization approach. They consider continuous integration as an important testbed for bug localization. That is why they motivate their study by describing several scenarios of bugs occurring in a project using continuous integration based on multiple discussions with software engineers. Laghari et al.\ \cite{ItemSetMining} note that integration tests provide a good context for bug localization research. Bugs in these tests occur rarely, however, when they do, both their complexity and impact are significant. Furthermore, finding such bugs can take hours and often has the largest priority, resulting in all other work being postponed until the bug has been resolved.

\subsubsection{Methodology}
The Patterned Spectrum Analysis beings by collecting traces for every test case. The trace contains data about the calls made to other methods whenever a test invokes a method from the base code. This data consists of a\textit{ caller object id} (object calling the method), \textit{caller id} (method from which the call is made) and a \textit{callee id} (called method). Subsequently, the trace is sliced into individual method traces for each executed method in the test case. The individual traces are processed by a closed itemset algorithm and turned into closed itemsets referred to as call patterns. A suspiciousness score is calculated and allocated to each call pattern by using a test coverage matrix and the call patterns of each method. This is done by using a fault locator (cf.\  Section 2). Finally, the suspiciousness of an individual method is the maximum suspiciousness of its call patterns.

\subsubsection{Evaluation}
\label{Itemset:eval}
Laghari et al.\ \cite{ItemSetMining} conduct experiments on 351 bugs from a subset of the dataset Defects4j (cf.\  Section \ref{evaluation:datasets}). The obtained results are compared against those of spectrum based bug localization (cf.\  Section \ref{spectrumb:based}) referred to by Laghari et al.\ \cite{ItemSetMining} as raw spectrum analysis. The raw spectrum analysis in the experiments uses the Ochiai fault locator \cite{ochiai}. The subset used for evaluation consists of Apache Commons Math, Apache Commons Lang, Joda-Time, JFreeChart, and Google Closure Compiler. Although the Defects4j dataset does not differentiate projects in terms of unit and integration tests, Laghari et al.\ \cite{ItemSetMining} perform an experiment and provide circumstantial evidence that the tests in the Google Closure Compiler are close to integration tests. This is important because Laghari et al.\ \cite{ItemSetMining} assume that integration tests execute multiple methods from different classes. Therefore, the spectrum analysis will contain more traces. In the cases of unit testing, the traces will be fewer.

Laghari et al.\ \cite{ItemSetMining} use \textit{wasted effort} as an evaluation metric which represents the number of results (methods) in the ranked list needed to be examined before getting to the method causing the bug.

$$\text{\textit{wasted effort}}=\frac{m +(n+1)}{2}$$

In the equation, $m$ represents the number of methods without bugs ranked strictly higher than the bug causing method and $n$ represents the methods without bugs with a rank equal to the one of the bug causing method.

The first observation is that patterned spectrum analysis results in less wasted effort than that of raw spectrum analysis. The strengths of the patterned spectrum analysis are especially visible when observing a certain bug in the Closure project, originating from a method with a unique call pattern in all failing test cases. This method is easily picked up by the patterned spectrum analysis, resulting in wasted effort of $0.5$ (i.e.\  the method is ranked highest in the result list). The wasted effort for the same bug when using raw spectrum analysis is $183$.

When evaluating how often is the wasted effort of both patterned and raw spectrum analysis $\leq 10$ (i.e.how many bugs are localized in the Top 10), Laghari et al.\ \cite{ItemSetMining} note that although the wasted effort of patterned spectrum analysis is lower than that of the raw one, a significant part of bugs is not contained in the Top 10.

The final evaluation investigates what is the effect of the number of triggered methods on the wasted effort. Interestingly, Laghari et al.\ \cite{ItemSetMining} observe that the number of triggered methods has a significant impact on raw spectrum analysis and almost none on patterned spectrum analysis.

\subsubsection{Advantages and Disadvantages}
In terms of wasted effort, the proposed patterned spectrum analysis provides better results than those of the raw spectrum analysis on the 351 bugs from the dataset. In addition, the patterned spectrum analysis ranks more bug causes in the  Top 10 ranked results. Furthermore, Laghari et al.\ \cite{ItemSetMining} hypothesise that patterned spectrum analysis performs significantly better than other spectrum based bug localization techniques when integration tests or tests of a similar structure are present, as in the case of the Google Closer Compiler. However, an inherent disadvantage of this approach is that methods that do not execute calls will always be placed at the bottom of the ranking. For example, this may be the case when a bug originates in a constructor.

\subsection{PredFL \cite{PredFL}}
There exist two groups of approaches used for bug localization that are similar. In particular, spectrum based \cite{ItemSetMining,PRFL} and statistical debugging based \cite{Liblit} approaches. Spectrum based methods (cf.\  Section \ref{spectrumb:based}) collect information about the coverage of program elements and use a formula referred to as a fault locator to calculate how suspicious each element is for causing a bug. Statistical debugging approaches seed predicates into the program and collect information about their coverage and their values. Based on this information, the importance of each predicate is calculated. While both types of methods use coverage information, they have been researched independently. Research on spectrum based approaches is focused on developing different fault locators (i.e.\  suspiciousness formulas), while work on statistical debugging is concentrated on different classes of predicates. Jiang et al.\ \cite{PredFL} propose Predicate-based Fault Localization (PredFL) \cite{PredFL}, an approach that combines both groups.

\subsubsection{Methodology}
To combine both groups of approaches Jiang et al.\ \cite{PredFL} consider Spectrum based fault localization (SBFL) as a Statistical debugging (SD) predicate. Because in Spectrum based localization the coverage of an element is important, this behaviour is easily translated to a predicate which evaluates to true when the element is covered. Moreover, SBFL values such as the number of successful/failed tests covering an element are mapped to the number of successful/failed executions in which a predicate is evaluated to true at least once. However, statistical debugging approaches and Spectrum based ones have different outputs. While the former results in a list of important predicates, the latter returns a list of suspicious elements. Therefore, Jiang et al.\ \cite{PredFL} decide that their unified model should return a list of suspicious elements as well. This is done so with the assumption that there exists a higher-order function called the combining method which computes the suspiciousness of an element by aggregating the importance scores of predicates. The final model consists of four parameters: a seeding function, a risk evaluation formula, a granularity function and a combining method.

Jiang et al.\ \cite{PredFL} perform several experiments in order to analyze the importance of the four aforementioned parameters. As a result of these experiments, they propose PredFL which is the unified model with a default configuration.

\subsubsection{Evaluation}
Because PredFL combines two groups of approaches and therefore is implicitly covered by existing techniques Jiang et al.\ \cite{PredFL} are interested in whether the approach is complementary to state-of-the-art techniques. They do so by integrating PredFL into CombineFL \cite{CombineFL}, a fault localization framework that consists of various types of techniques, such as spectrum based, information retrieval based, etc. All of these techniques are distributed in levels according to their execution time, from seconds (Level 1) to hours (Level 4). After integrating PredFL into CombineFL, their combined performance is evaluated on the Defects4j dataset (cf.\ Section \ref{evaluation:datasets}) using the Recall at Top 1,3,5,10 metric (cf.\ Section \ref{evaluation:metrics}). Note that for the experiments mentioned in the Methodology section the same dataset and evaluation metric was used with the addition of the EXAM score (cf.\ Section \ref{evaluation:metrics}).

Jiang et al.\ \cite{PredFL} observe that after the integration of PredFL, CombineFL improves its recall at all Top levels (1,3,5,10) with 4-8\%. Thus, they conclude that PredFL is complementary to existing techniques.

\subsubsection{Advantages and Disadvantages}
PredFL combines two types of approaches successfully and as shown by Jiang et al.\ \cite{PredFL} can be used to improve the performance of another technique. However, there exist several concerns. Firstly, the technique is not evaluated as a standalone against the performance of state-of-the-art SBFL and SD approaches. Therefore, it is not sufficiently investigated whether the benefit of combining both types of techniques is significant or not. Secondly, the implementation made available by Jiang et al.\ \cite{PredFL} is built on Java Development Tools (JDT), used for the generation of predicates. This makes the implementation language-specific. Furthermore, if the strategy of PredFL is to be implemented in another language, a way to generate and seed predicates into the program is needed. This may be difficult or not possible and could impact performance.

\subsection{PRFL \cite{PRFL}}
There exist various bug localization techniques and tools based on Spectrum analysis. Zhang et al.\ \cite{PRFL} note that although there is a multitude of techniques, there isn't one that is the best performing. They attribute this to the fact that all the methods focus on differentiating the programming entities (e.g.\ methods, executable statements, etc.) which represent one aspect of program spectra. There is no work concentrating on the differentiation of tests, which represent another aspect of program spectra. Zhang et al.\ \cite{PRFL} explore the contribution of different tests to improve on the weaknesses of existing spectrum based techniques. They present PRFL \cite{PredFL}, a method based on PageRank that improves Spectrum based bug localization by taking into consideration additional test information with the use of the PageRank algorithm \cite{page1999pagerank}. After PageRank is used to recompute spectrum information existing spectrum based (fault locator) formulas can be utilized, resulting in better bug localization performance.

PageRank \cite{page1999pagerank} is an algorithm that improves the speed and quality of a search. In the context of a web search, PageRank considers the World Wide Web as a graph containing nodes (i.e. web pages) that are linked. Intuitively, the algorithm gives higher importance to nodes that are linked by important nodes and lower importance to those linked by unimportant nodes. PageRank has been applied in different domains such as Biology, Chemistry, Neuroscience, Recommendation systems, Social networks and others.

\subsubsection{Methodology}
PRFL consists of three main phases. The first phase is the preparation phase. This phase uses both static and dynamic analysis to obtain test coverage information and construct a call graph that maps the runtime connections of methods. This is followed by the second phase - the PageRank analysis. In the second phase, weighted spectrum information is generated for all the tests. In the final phase, referred to as the Ranking phase, fault locators (i.e. Spectrum based equations) are applied on the weighted spectrum to rank methods in the source code. The weighted spectrum does not only include information from the test coverage, but also information from the call graph constructed in the first phase as well as test scopes. Zhang et al.\ \cite{PRFL} argue that the weighted spectrum reflects more accurately whether a method is faulty or not, therefore, improving the efficiency of spectrum based techniques.

\subsubsection{Evaluation}
Zhang et al.\ \cite{PRFL} two datasets for the evaluation of PRFL. The first one contains real-world bugs and consists of five projects containing 357 bugs (JFreeChart, Closure Compiler, Apache Commons Lang, Apache Commons Math, Joda-Time) from the Defects4j dataset (cf.\ Section \ref{evaluation:datasets}). The second dataset consists of artificially created bugs. Zhang et al.\ \cite{PRFL} argue that the real-world bugs are somewhat limited. For that reason, they use the PIT mutation testing tool \cite{PIT} to create mutation bugs using 87 of the most popular Java projects on Github. This results in 30 692 artificial bugs. The evaluation metrics used are AWE (absolute wasted time), which is the same as the wasted time metric described in Section \ref{Itemset:eval}, and Recall at Top 1,3,5 (cf.\ Section \ref{evaluation:metrics})

Initially, in the evaluation PRFL is compared against state-of-the-art spectrum based techniques. This is done by recording the performance of each technique without the use of PRFL and then the performance of the technique combined with PRFL on all the projects from the Defects4j dataset. The versions of the techniques containing PRFL outperform their counterparts. Interestingly, PRFL not only improves AWE and the Recall at Top N, but it provides the biggest improvement when combined with the best performing techniques. In addition, Zhang et al.\ \cite{PRFL} evaluate the overhead brought by the computation of the call graph and the page rank analysis. They find that the overhead is insignificant (in the order of seconds) supporting the claim that PRFL is a lightweight technique.

Next, Zhang et al.\ \cite{PRFL} study the impact of the number of bugs on PRFL. Defects4j contains single-faults and multi-faults. In the single-fault, a single bug is present in the program. In the multi-fault, there can be multiple existing bugs. Zhang et al.\ \cite{PRFL} split the previous results into results on single-fault and results on multi-fault. They observe that state-of-the-art techniques perform inconsistently on the two sets. For example, the Op2 technique \cite{naish2011model} which outperforms all other techniques on single-faults, performs the worst on multi-faults. Although PRFL boosts the performance of all techniques in both cases, the improvements are larger on single-faults.

Finally, PRFL is evaluated on artificial bugs. This is performed in a similar manner as the evaluation on real bugs. The evaluation shows that the type of bug (i.e.\ real/artificial) does not impact the performance of PRFL.

\subsubsection{Advantages and Disadvantages}
PRFL is advantageous because it boosts the performance of state-of-the-art spectrum based techniques while having an insignificant overhead. Nevertheless, as Zhang et al.\ \cite{PRFL} express in the introduction of their work, the performance of state-of-the-art spectrum based techniques is limited and there is not a best performing one. Taking this into consideration, Zhang et al.\ \cite{PRFL} do not establish whether the boost of PRFL is significant enough that the technique can be used in practice. Furthermore, the results of spectrum based techniques are worsened in the presence of multiple bugs (multiple-fault case). Unfortunately, in practice, this case may commonly occur.

\subsection{Recommendation}
Every tool and technique investigated in this study is unique. Not only because of the difference between the types of approaches (e.g. Information Retrieval and Spectrum Based), but also methods sharing the same type are also distinct. For example, while both BugLocator \cite{BL} and BLUiR \cite{BLUiR} are information retrieval based, both handle the challenge of bug localization differently. Although each tool has different benefits and disadvantages and its performance might fluctuate in different contexts, I find Legion \cite{Legion} to be the most beneficial among all considered tools and techniques. This subsection contains the reasoning behind this choice.

Firstly, information retrieval is among the most widely used strategies for bug localization. A lot of research has been done and is being done that boosts the performance of information retrieval techniques and tools for bug localization. Although IR based bug localization methods struggle with the use of synonymous words in bug reports and source code files, IR approaches can be combined with ML and DL techniques to alleviate this problem. In addition, as displayed by Legion, the inclusion of additional corpora can also improve the associations between bug reports and source code.

Legion is an extension of a state-of-the-art IR based bug localization approach, built and evaluated in an industrial setting. While many tools and techniques attempt to use real-world bugs, Legion's performance is evaluated on closed source repositories used in industry. Furthermore, Legion is among the very few tools evaluated on an unconventional metric - the expectations of developers. The majority of other new studies do not consider such expectations. Researchers consider mainly the metrics employed by previous state-of-the-art techniques and attempt to outperform them. Even though this evaluation is important, it does not aid the adoption of bug localization techniques by software practitioners. Legion, however, displays performance sufficient for use by junior and less experience developers.

Finally, Jarman et al.\ \cite{Legion} make several propositions for future work that could improve Legion even further.

\section{Conclusion}
\label{conclusion}
This study investigated different bug localization tools and techniques of different types. The focus of the work was on tools and techniques evaluated in an industrial setting. However, due to the lack of such methods, works evaluated on open source projects were also included. Each work was briefly introduced followed by a description of its evaluation, as well as its advantages and disadvantages. Finally, a recommendation was presented in which Legion \cite{Legion} was proposed as the most advantageous approach among all considered in this study for use in the industry.

\section*{Acknowledgment}
I am grateful to Dr Eleni Constantinou (Eindhoven University of Technology) and Dr Dennis Dams (ESI-TNO) for providing scientific guidance.

\bibliographystyle{ieeetr}
\bibliography{review.bib}
\end{document}